
\input harvmac

\Title{hep-th/9406132 RU-50-94 SCIPP 94/12}
{\vbox{\centerline{Coping With Strongly Coupled String Theory}}}
\bigskip
\centerline{Tom Banks}
\smallskip
\centerline{\it Department of Physics and Astronomy}
\centerline{\it Rutgers University, Piscataway, NJ 08855-0849}
\smallskip
\centerline{Michael Dine}
\smallskip
\centerline{\it Santa Cruz Institute for Particle Physics}
\centerline{\it University of California, Santa Cruz, CA   95064}
\bigskip
\baselineskip 18pt
\noindent
String theory, if it describes nature,
is probably strongly coupled.  As a result,
one might despair of making any statements about the
theory.  In the framework of
a set of clearly spelled out assumptions, we show
that this is not necessarily the case.  Certain discrete gauge
symmetries, combined with supersymmetry, tightly
constrain the form of the effective action.  Among our assumptions
are that the true ground state can be obtained from
some perturbative ground state by varying the coupling, and
that the actual numerical value of the low energy field theoretic
coupling ${g^2 \over 4\pi}$ is small.  It follows that the low energy
theory is approximately supersymmetric;
corrections to the superpotential
and gauge coupling function are small, while corrections to
the Kahler potential are large; the spectrum of light particles
is the same at strong as at weak coupling.  We survey
the phenomenological consequences of this viewpoint.
We also note that the
string axion can serve as QCD axion in this framework (modulo
cosmological problems).

\Date{5/94}

\newsec{\bf Introduction:  Can One Make a Sensible
Superstring Phenomenology?}

Weakly coupled string theory is a phenomenological disaster.  In some of
its classical ground states it contains a spectrum of particles
tantalizingly reminiscent of the world as we know it.  But in addition
these ground states have a large spectrum of unwanted massless
particles, generically called moduli\foot{We include under this rubric
the dilaton and its superpartners.}.
Their presence indicates that perturbative string theory is
grossly inconsistent with observation.  They contradict the weak
equivalence principle, and are thus in conflict with the Eotvos-Dicke
experiment.  They lead to time variation of the fundamental constants
that is in contradiction with observation, and they predict unobserved
perturbations of the motion of the planets.

The oriental screen behind which
string theorists hide this embarrassment is called
{\it nonperturbative physics}.  After all, while string theory predicts
the existence of the quarks, leptons and gauge bosons of the Standard
Model, perturbative string theory predicts that they are all massless.
Surely it is plausible that the same nonperturbative mechanism which
produces the observed mass spectrum, will rid us of the embarrassing
moduli.  This plausible sounding excuse runs afoul of some special properties
of string theory first pointed out by Seiberg and one of us\ref\dineseiberg{
M. Dine and N. Seiberg,
Phys. Lett. {\bf 162B}, 299 (1985),
and in {\it
Unified String Theories}, M. Green and D. Gross, Eds. (World Scientific,
1986).
}.
String theory has, to our knowledge no free parameters apart from a
fundamental length scale.  If it is weakly coupled, this is only because
the vacuum expectation value (VEV) of the dilaton takes on a special
value. But this value is dynamically determined by the effective
potential of the dilaton, which itself should be computable in a
systematic asymptotic expansion in the coupling.  We know its value,
namely zero, in the extreme weak coupling limit.  It will approach zero
according to some well defined asymptotic formula, which is either
positive and monotone decreasing, or negative and monotone increasing as
the coupling goes to
zero\foot{Mathematically, the montonic behavior could be modulated
by some sort of oscillation.  We know of no physical mechanism which
could produce such an oscillation.  In any event, oscillatory modification
of a monotonic function could at best produce an infinite number
of false vacua for the dilaton.}.  In neither case can the potential have a
minimum for parametrically weak coupling.  Almost by definition, a
nontrivial minimum of the dilaton potential implies that terms of
different order of the asymptotic expansion must make comparable
contributions.  Thus, there are at least some effects in string theory
which do not admit a controllable weak coupling expansion.  In view of
this we should be suprised if anything were to
be calculable in such an expansion.  If string theory is sufficiently
strongly coupled to stabilize the dilaton, why should we believe any
weak coupling calculation in the theory?

There are many possible responses to this situation.  Perhaps the
most reasonable is to discard the theory altogether.  Still,
given its many attractive features, particularly the fact
that it is our only theory of quantum gravity, it is hard to
resist the temptation to look for other ways out.  Among these,
we can hypothesize that some
presently unknown modification of the theory will eliminate the dilaton
but preserve the more attractive aspects of string theory.  We can hope
that group theoretical factors conspire to allow two terms of different
order in the weak coupling expansion to be of comparable orders of
magnitude even when the coupling is weak.  This is the philosophy behind
the so called racetrack models\ref\race{V. Krasnikov,
Phys. Lett. {\bf 193B} (1987) 37;
L. Dixon, in the {\it Proceedings of the DPF Meeting,
Houston, 1990}; J.A. Casas, Z.
Lalak, C. Munoz and G.G. Ross,
Nucl. Phys. {\bf B347} (1990) 243;
T. Taylor, Phys. Lett. {\bf B252} (1990) 59.}, in which factors of the form
$e^{-{8\pi^2 \over N g^2}}$ and $e^{-{8\pi^2 \over (N+1) g^2}}$ for
large $N$ compete to give a minimum of the potential.
Finally, we can bite the bullet, and admit that string theory is
strongly coupled.

Is there any utility to such an admission or does it simply tell us that
the dynamics of string theory is at present incomprehensible?  Is the
theory's
hypothetical applicability to the real world destined to remain a
hypothesis until we learn how to solve the strongly coupled problem?
We would like to argue in the present paper that the answers to these
questions are negative.  The situation is not completely without
precedent. The historical development of condensed matter physics
depended entirely upon the fact that, although the fundamental theory of
electrons interacting via Coulomb potentials contains no small
parameter, the low energy dynamics of this system is, in many regimes,
dominated by a set of weakly coupled excitations with the quantum
numbers of the fundamental electrons.  As a first step in coping with
strongly coupled string theory we make the analogous {\it hypothesis}:
even though the fundamental short distance degrees of freedom in string
theory are strongly coupled, the low lying spectrum of the full solution
of the theory has the same quantum numbers and multiplicities as the
massless spectrum in one of the theory's myriad classical ground states.
To put it another way: string theory certainly has a large number of
metastable states concentrated in the region of field space where the
entire theory is weakly coupled.  We assume, that as we move in to the
strong coupling region, the low lying spectrum of at least one of these
classical vacua becomes the true spectrum of the strongly coupled
theory\foot{A number of authors have conjectured the existence of a weak
to strong coupling duality transformation in string
theory\ref\sduality{A. Font,
L.E. Ibanez, D. Lust and F. Quevedo,
Phys. Lett. {\bf B249} (1990) 35;
S.J. Rey, Phys. Rev. {\bf D43} (19911) 526; A. Sen,
Nucl. Phys. {\bf B404} (1993) 109; Phys.
Lett. {\bf B303} (1993) 22; Mod. Phys. Lett.
{\bf A8} (1993) 2023; J. H. Schwarz
and A. Sen, Nucl. Phys. {\bf B411} (1994) 35.}.
The infinitely strongly coupled theory based on
one classical ground state is equivalent to a weakly coupled theory
based on another.  In the context of this conjecture we should replace
the phrase ``strong coupling'' in our discussion by ``intermediate
coupling''.  We are talking about a regime in which, at short distances,
there is no  description of the theory in terms of weakly coupled
semiclassical excitations.}.  We will see below that there are some
plausible pieces of evidence for this assumption.

The question of which of the classical ground states determines the true
spectrum may ultimately be answered only by strong coupling physics.
Here we pursue a more modest goal. We assume a particular ground state
and try to find constraints on low energy
physics in this ground state which will be valid independently of
the details of strong coupling physics.   We find that many, but by no
means all, of the
predictions of perturbative string theory can be viewed as consequences
of certain discrete gauge symmetries, and are reproduced even when the
coupling is strong.  The discussion of these symmetries and their
consequences is the contents of section II.

In section III we take up the question of how the low energy excitations
of string theory can be weakly coupled when string theory is strongly
coupled.  The situation is not quite analogous to condensed matter
physics, because the infrared behavior of Yang Mills theory is strongly
coupled. Thus, one must explain why the gauge and Yukawa coupling
parameters in the effective Lagrangian just below the string scale are
weak.  We identify two possible explanations of this fact.  The first
involves the notion that weak coupling means something quantitatively
different in string theory than it does in field theory.  String
perturbation series are more divergent than field theoretic
series\ref\shenker{S.H. Shenker, talk given at the
Cargese Workshop on Random surfaces, Quantum Gravity
and Strings, Cargese, France (1990), Published in
the Proceedings, and private communication.}.
Correspondingly, we expect nonperturbative
corrections to be large when functions of the coupling like
$g^{-p}e^{-b/g}$ are of order one.  Here $b$ and $p$ are real constants
($b>0$) which we do not know how to compute for realistic string
theories.  On the other hand, nonperturbative effects in the low energy
effective field theories derived from string theory are generically of
order $g^{-k} e^{- {8\pi^2 \over N g^2}}$, where $N$ and $k$ are
positive constants of order $10$ or less.  These effects are tiny at
the putative value of the unified gauge coupling\ref\susyunif{
U. Amaldi, W. de Boer and H. Furstenau, Phys. Lett.
{\bf B260} (1991); U. Amaldi {\it et al.}, Phys. Lett. {\bf B281} (274) 1992.}
${g^2 \over 4\pi} \sim 1/25$, but it is perfectly plausible that the
stringy contribution is of order one at this value of the coupling.

One might also attempt to understand the
discrepancy between the string and four
dimensional field theoretic couplings in terms of the volume of the
compactified dimensions.  Conventional wisdom in weak coupling string
theory is that the scale of these dimensions must be the same as the
string scale\ref\vadim{V.S. Kaplunovsky,
Phys. Rev. Lett. {\bf 55} (1036) 1985.}\ref\ds{M. Dine
and N. Seiberg, Phys. Rev. Lett. {\bf 55} (366) 1985.}.
We explain how this constraint may be
relaxed in a strongly coupled theory.  A large compactification volume
may also help to explain the difference between the string scale and the
``observed'' scale of coupling unification\susyunif, without recourse to large
threshold corrections.  This idea is very tightly
constrained by the ``observed''
values of the unified coupling and unification scale.  We argue that
no matter how strong the coupling in the higher dimensional theory,
a Kaluza-Klein ansatz with more than one dimension as large as the
``observed'' unification scale, would lead to an unacceptably small unified
coupling.  A Kaluza-Klein vacuum with one large internal dimension is
acceptable on purely numerical grounds. However,
the dilaton couplings in such a theory are
highly constrained by the combination of approximate $5$ dimensional
SUSY and discrete shift symmetries.  This may make it impossible to carry
out our program for stabilization of the dilaton in such a theory.

Having made the strong coupling string theory/weak coupling field theory
dichotomy plausible, we explore how these ideas illuminate the central
problems of stabilization of the dilaton and supersymmetry breaking.
We argue that a particularly attractive resolution of these problems may
result from the interaction of a nonperturbatively determined Kahler
potential for the dilaton and other moduli, and a single gaugino
condensate.  In weakly coupled string theory, a single gaugino
condensate leads to a runaway vacuum, but the nontrivial Kahler
potential may stabilize it at strong coupling.  An essential feature of this
mechanism
is that discrete symmetries constrain the form of the superpotential to be that
determined by the low energy gaugino condensate.  Stringy nonperturbative
corrections to this are very small.  The Kahler potential's dependence on the
real
part of the dilaton-axion superfield
is completely unconstrained by the symmetries (they involve shifts
of the axion only), and feels the full force of nonperturbative stringy
physics.
We try to outline the low energy phenomenology
which can be predicted in such a model.
When this sort of model for SUSY breaking is combined with the discrete
gauge symmetries that we have imposed to preserve predictability, one
sometimes\foot{{\it i.e.}
for some discrete gauge groups}
finds that the dominant contribution to the mass of the model independent
axion comes from nonperturbative QCD dynamics.  Consequently, it can
solve the strong CP problem, but the model may predict a
cosmology at variance with observations.

To summarize, we have tried to face
squarely the problem of strongly coupled
string theory, and found that it is not as hopeless as one might
imagine.  The assumptions which are required (e.g.
vanishing of the cosmological constant at the minimum
of the potential) are no stronger than those required for
any string phenomenology.  One gluino condensate
serves to both break supersymmetry and stabilize the
dilaton, and resolves the Dine-Seiberg
problem\foot{Apart from the cosmological
version of the problem discovered in \ref\bruhart{R. Brustein
and P.J. Steinhardt, Phys. Lett. {\bf B302} (1993) 191.}.  This will be dealt
with in  \ref\modcosm{
T. Banks, M.Berkooz, G. Moore, S. Shenker, P.J.Steinhardt, to appear.}}.
Good predictions of perturbative
string theory, such as the form of the spectrum, are
preserved.
Indeed the detailed computation of Yukawa couplings, possible
in the perturbative approach and impossible in ours, always suffered from
the problem that one did not know which weakly coupled minimum was the
true ground state.
Alternatively, a Kaluza-Klein
scenario might provide an {\it explanation}
of the weakness of high energy field theoretic couplings (as well as a
simple reconcilation of string theory with the ``observed'' scale of
coupling unification) in terms of the volume of the compactified
internal dimensions of space.  However, it may be impossible
to achieve a simple stabilization of the dilaton in such a theory.

\newsec{Constraints on Non-Perturbative Physics from Discrete
Symmetries and Their Implications for Strong Coupling}

The tool which we will use throughout this paper is discrete gauge
symmetry.  Typical classical string vacua manifest a plethora of
discrete symmetries.  In the large radius limit for the internal
nonlinear model, many of these
symmetries can be seen to be general coordinate transformations of the
internal space.  Other, peculiarly stringy, symmetries like duality
can also be viewed as gauge symmetries by finding points in moduli space
where they become incorporated in low energy continuous gauge
groups\ref\huet{M. Dine, N. Seiberg
and P. Huet, Nucl. Phys. {\bf B322} (1989) 301.}.
It is tempting to speculate\huet\
that all discrete symmetries of string theory are gauge symmetries, and
should therefore be preserved by any perturbative or nonperturbative
effects in the theory.  To date, all apparent anomalies\ref\dineetal{
T. Banks and M. Dine, Phys. Rev. {\bf D45} (1992)
424; L.Ibanez and G.G. Ross, Phys. Lett.
{\bf  260B} (1991) 291; Nucl. Phys. {\bf 368} (1992) 3.}
that have been discovered in these transformations can be cancelled by a
discrete analog of the Green-Schwarz mechanism.  Beyond the tree
approximation, the dilaton superfield transforms nontrivially under the
symmetry in order to cancel an apparent anomaly in fermion
transformation laws.

Before applying these symmetries to strongly coupled string theory we face
two barriers which seem to prevent their efficient application.  The
first is a technical problem involving field definitions.  It is related
to the notorious ``multiplet of anomalies'' problem which has haunted
supersymmetric gauge theories for years.  We will define the problem and
deal with it in the subsection immediately below.  The second barrier to
the use of symmetries of a classical vacuum in a strongly coupled theory
is spontaneous symmetry breakdown.  How can we tell that the strongly
coupled theory is not in a different phase from the classical one? Examples of
such {\it phase transitions} abound in field theory. To mention but one: the
$Z_2$ symmetry of the dual variables in the low temperature two
dimensional Ising model is spontaneously broken in the strong coupling
region. This is potentially
a serious problem as we pass
from weak to strong coupling in string theory,
but once again, the combination
of supersymmetry and discrete
symmetries comes to our rescue and forbids such transitions.
We will present our argument in the
second subsection below, and then proceed
to apply discrete symmetries to predict properties of strongly coupled
string theory.

\subsec{In Which the Conventions Are Observed}

The bosonic component of the dilaton superfield is conventionally
defined to be $S = {8\pi^2\over g^2} + ia$ where, in classical string
theory, $g$ is the string coupling and $a$ the dimensionless axion field
(dimensions are supplied by the string tension $\alpha^{\prime}$).  Shortly,
we will present evidence
which suggests that
physics is periodic in the axion, with period $2\pi$.  This periodicity is an
example of the kind of discrete gauge symmetry that we will be invoking.
The other
discrete symmetries we will discuss shortly are
gauge and general coordinate transformations in some internal
space, and thus are definitely genuine discrete gauge symmetries.
For the symmetries of interest to us, the
model independent axion must have a non trivial transformation law in
order to cancel anomalies (a discrete version of the Green-Schwarz
mechanism).  These transformations involve axion
shifts by fractional multiples of $2 \pi$.  $2\pi$ periodicity of the
axion is not related to continuous gauge symmetries in an analogous
manner. However, if it is a valid symmetry of string theory then it
certainly shares the major property of discrete gauge symmetries in
that it will not be violated by wormholes.

When the axion is coupled to low energy supersymmetric gauge theories in
the conventional way, a tension develops between the desire to have the
real part of the $S$ field be related to the coupling in some particular
regularization
scheme while the imaginary part still transforms properly under symmetry
transformations. This is related to the multiplet of anomalies puzzle:
the stress tensor is in a supermultiplet with an axial current, whose
divergence can apparently be computed exactly at one loop.  One would
then expect the trace of the stress tensor, and thus the $\beta$
function, to vanish beyond one loop.  Of course it doesn't,
in conventional renormalization schemes.

This problem was essentially solved many years ago by Shifman et.
al.\ref\shifman{M.A. Shifman and A.I. Vainshtein,
Nucl. Phys. {\bf B277} (1986) 456 and
Nucl. Phys.. {\bf B359} (1991) 571.}.
These authors observed that the paradox could be
resolved by choice of a special scheme for coupling constant
renormalization and for the normalization of the axial current.
Supersymmetry and the Adler Bardeen theorem (in the guise of an exact
instanton computation) enabled them to compute the exact $\beta$
function in their scheme.  We will add a small twist to their procedure,
which is useful for our purposes.

We will use a definition of the coupling constant which preserves its relation
to the axion field which transforms simply under various symmetries of
the theory.
These symmetries all act by shifts of the axion by discrete amounts.
${8\pi^2\over g^2}$ is defined to be the superpartner of this axion field, in
the sense
that $S \equiv {8\pi^2\over g^2} +
i {a}$ is the $A$ component
of a chiral superfield.  This superfield is related to that defined by the
coupling constant, $S_c$,  in a ``conventional"
regularization scheme (one which preserves the universality of the two loop
beta function) by a nonanalytic transformation of the form
$S = S_{c} - {b_1 \over b_0} \ln [S_{c}] + \ldots$,
where $b_0$ and $b_1$ are the first two coefficients in the beta function.
Although nonanalytic at $S = \infty$, this tranformation is locally analytic,
and preserves SUSY.  We prefer this definition because it
dramatically simplifies the formulae for the nonperturbative contributions
to the superpotential and gauge kinetic term.  All complications are shifted
into the Kahler potential, which will be uncomputable anyway in our framework.

\subsec{In Which Phases Defend Against Phase Transitions}

We now come to what is probably the most important point of our analysis.  We
would like to use various
discrete symmetries of perturbative string vacua to constrain the
nonperturbative behavior
of the theory.
\foot{Some preliminary steps in this direction were taken in
\ref\dine{M. Dine
SCIPP 93-30 (1993), to appear in Proceedings
of STRINGS 93.}.}
We intend to study the effective lagrangian of the theory at a
scale below the string scale, but above the scale of any strong nonperturbative
field theoretic behavior, and we wish to claim that this lagrangian is
invariant
under the anomaly free discrete symmetries of the perturbative ground state
{\it even when nonperturbative
effects due to massive string modes are taken into account}.
We will assume that these symmetries are not
broken explicitly; needless to say, lacking a non-perturbative
definition of the theory, we cannot say anything
rigorous about this question.   But we can show that the
assumption that anomaly free symmetries
remain unbroken nonperturbatively is built into all considerations
of string theory.
Let us consider what happens to one of our anomaly free symmetries
when we move about on the moduli space of classical string ground states,
following a path along which
the symmetry remains perturbatively unbroken.  In all cases\foot{Apart
from the $2\pi$ shift of the axion.} of which we are aware, one can connect
the ground state continuously to flat ten dimensional space.  In this limit our
symmetry becomes
a ten dimensional Lorentz transformation or gauge transformation, and the axion
shift which must accompany
the symmetry tranformation for purposes of anomaly cancellation, is
seen to be
a special case of the Green-Schwarz mechanism.  We believe that this implies
that
an explicit nonperturbative violation of the discrete symmetries we are
discussing
would have the same status as a violation of local Lorentz invariance.
Perhaps nonperturbative string theory does not preserve local Lorentz
invariance,
but if so, one must fear for the consistency of the theory.

Notice that this argument does not apply to the {\it continuous} axion shift
symmetry of perturbative
string theory.  This symmetry is explicitly broken to a discrete subgroup
by low energy gauge theory instantons.  At the level of renomalizable
interactions,
it is sometimes possible to combine the axion shift with continuous global
symmetries
of the low energy gauge action to construct an anomaly free $U(1)$.  However,
general theorems in
string theory\ref\banksdixon{T. Banks,
L. Dixon, D. Friedan and
E. Martinec, Nucl. Phys.
{\bf B299} (1988) 613.}
assure us that these continuous global symmetries are {\it accidental}.
They are broken to discrete subgroups by higher dimension terms in the action.
Thus we expect that
nonperturbative effects of high energy string modes will also break the axion
shift symmetry to a discrete
subgroup.

While it is not in any obvious way connected to local Lorentz
or gauge invariance, we believe that the discrete subgroup
of $2\pi$ shifts of the
axion {\it is} an exact symmetry of string theory.
The key argument for nonpertubative validity of the
discrete axion shift
symmetry is based on the notion that string
instantons can be
regarded as conformal field theories.  We will also need to recall the
quantization of the three-index antisymmetric tensor
discussed by Rohm and Witten.\ref\rohmwitten{R. Rohm
and E. Witten, Annals of Physics {\bf 170}
(1986) 454. }  We imagine compactifying
ordinary four-dimensional space-time on some large surface.
Then the quantization condition is the statement that the integral
\eqn\hquantization{\int d^3 \Sigma ~H = n.}
Now consider an Euclidean conformal field theory corresponding to
 some localized field configuration
(i.e. some configuration involving massive string fields).
At large distances, the world sheet lagrangian approaches that of a
weakly coupled nonlinear model with an axion field which behaves as
\eqn\axionfalloff{a \sim {n \over r^2}.}
The change in the Peccei-Quinn charge is related
to the axion field by
\eqn\pqchange{\Delta Q_{pq}= \int d^4 x~\partial^2 a.}
On the other hand, the axion and $h$ are related by
\eqn\axionh{h = da.}
Substituting in the quantization condition \hquantization, we learn
that the change in the Peccei-Quinn
charge is also $n$.  This is precisely the change we would have obtained
from ordinary gauge theory theory instantons.  This argument
suggests that only operators of the form $e^{ina}$ can
be generated by nonperturbative string physics.

The main limitation of this argument is that it is not clear
in what sense non-perturbative string physics is described by
two dimensional field theories.  Matrix models, for
example (or simply the analogy with QCD) suggest
that the relevant degrees of freedom to a non-perturbative
analysis might be different than those of
string perturbation theory.  No connection between ``instanton conformal
field theories'' and the nonperturbative physics in these models has
been established, and the relevance of the Rohm-Witten quantization
condition can be questioned. In what follows,
we will assume that this quantization is true non-perturbatively.
In particular, we will assume that terms like $e^{2 \pi ia/N }$,
which might otherwise be permitted by symmetries,
cannot appear in the effective lagrangian just below
the string scale.  We will comment briefly on the consequences of
relaxing this assumption.


Some readers may object
that gluino condensation generates superpotentials
which behave as $e^{ia/N}$, for some integer
$N$.  However, it is not hard to see that this
is consistent with the discrete symmetry.  Indeed,
the gluino condensate is proportional to
\eqn\condensatephase{e^{ia/N} e^{2\pi i n/N}}
reflecting the fact that the condensate spontaneously
breaks an (in general
approximate) $Z_N$ symmetry of the theory.
Thus a $2 \pi$ shift of $a$ can be compensated
by a change of the choice of branch in the condensate.
Indeed, if one formulates gluino condensation
along the lines of ref. \ref\veneziano{G. Veneziano
and S. Yankielowicz, Phys. Lett. {\bf 113B}
(1982) 231.} then the gluino condensate
is obtained by solving an equation of the form
\eqn\condensateroot{(\lambda \lambda)^N \propto e^{ia}}
which clearly respects the symmetry.  Thus, the discrete axion shift
symmetry appears to be an exact symmetry of string theory which is
spontaneously broken by gaugino condensation.

In view of the spontaneous breakdown of discrete gauge symmetries by the
strongly coupled gauge theory
in the gaugino condensate scenario, one is moved to worry about the possibility
that the strongly coupled
short distance degrees of freedom of string theory might also spontaneously
break these symmetries.  If this were to happen,
the symmetries would impose no constraints whatsoever on the low energy
effective
lagrangian\foot{It is appropriate to comment
here on the following puzzling question:  All of the discrete symmetries that
we employ are, in a sense,
spontaneously broken at a high scale because they are realized through the
nonlinear
transformation law of the model independent axion.
We have just noted that high energy breaking of discrete symmetries generally
leaves
no traces in the low energy action.  Where then do our constraints come from?
The special situation that is realized here is a consequence of the fact that
the axion appears in the low energy theory, because it can be viewed as the
pseudo-Goldstone boson of an approximate, accidental, continuous symmetry.
The approximate validity of this symmetry is a consequences of the dual
constraints of the discrete symmetries and supersymmetry.  Thus spontaneous
breakdown of the discrete symmetries through the axion can be seen explicitly
in the low energy lagrangian.  Some issues involved in the spontaneous breaking
of discrete symmetries of this kind are discussed in Appendix
A, where they are illustrated in Supersymmetric QCD.
The question we deal with below is whether there can be further spontaneous
breakdown of the discrete symmetries due to VEVs of massive fields.}.

Spontaneous breakdown of perturbative discrete symmetries by strongly
coupled short distance physics is more difficult to rule
out in  general,  but within the strong coupling framework
we have outlined, one can give a compelling argument
against it.
Let us begin by studying the extreme weak coupling region of moduli space,
where $|S|$ is large.
Remember that our fundamental assumption is that the true quantum mechanical
ground state of string theory can be reached by following a continuous path
from a point
in this region towards strong coupling
(in the sense discussed in the next section).
Without such an assumption we cannot even begin to discuss the
strong coupling region unless we know how to solve directly for the spectrum
there.  Of course, one might worry that the spectrum
of the theory changes as we move from weak to strong
coupling.  But as we will see below, this cannot occur.
We also assumed that at zero coupling (i.e. in the classical
string model) the theory is supersymmetric.  We will
see that as a consequence of this assumption, the
theory is approximately supersymmetric at strong
coupling as well (e.g. at low energies, it looks like
a supersymmetric theory with explicit soft breakings).
This means that even in the strong coupling framework,
SUSY can be related to the solution of the hierarchy problem.
This is not something one might have expected a {\it priori}.

 Returning then to the weak coupling region, we note that in this region,
integrating out the heavy string
modes cannot lead to spontaneous breakdown of discrete symmetries observed in
perturbation theory.
The heavy modes are weakly coupled.  Their classical vacuum expectation values
are zero, and finite action field configurations
must approach these VEVs at spatial infinity.  Thus, even when nonperturbative
effects are
taken into account, the discrete symmetries of perturbation theory are
preserved.
As we move into the strong coupling region, this argument breaks down.  In
ordinary bosonic field theory we could encounter
either a first or a second order phase transition at some finite value of the
coupling.


To examine the possibility of spontaneous breakdown
via the VEV of a heavy field
, we imagine including the zero modes of the heavy fields in the effective
superpotential\foot{Indeed, in a strict Wilsonian
approach, one should always keep the low momentum modes of all
fields in the effective action.  Typically, the low momentum modes
of fields with masses larger than the cutoff may be integrated out
classically even if the full theory has no small parameters.  This
is why one usually ignores them.}.  At weak coupling,
 the dynamics of the heavy fields does not break supersymmetry.  Thus,
the equation determining the VEVs of heavy fields is
\eqn\hevvev{\partial_{\Phi^i}(W_0 + \delta W) + {1\over
M_P^2}\partial_{\Phi^i}(K_0 + \delta K)(W_0 + \delta W) = 0}
Here $W_0$ and $K_0$ are the tree level superpotential and Kahler potential
respectively, while $\delta W$ and $\delta K$ are the
quantum corrections to them.  $\delta W$ receives only
nonperturbative corrections, while $\delta K$ has a perturbation
expansion.  The solution of the tree level equations
is $\Phi^i = 0$, and it is stable, in the sense that none of the $\Phi^i$
directions is flat.  Near $\Phi^i = 0$, $\partial_{\Phi^i}W_0
 + {1\over M_P^2}\partial_{\Phi^i}(K_0)W_0$ has the form $H_{ij}\Phi^j$, with a
nonsingular matrix $H$.

The corrections to the tree level equations coming from Kahler
potential
terms (and indeed, the tree level Kahler potential contribution
itself)
are all proportional to ${1\over M_P^2}$, while $H_{ij}\propto
M_P$.
Thus unless $\partial_{\Phi^i} \delta W$ is large we can solve these
equations perturbatively.  In that case, since the equations are
covariant under the discrete symmetries in question, the
expectation
values of each heavy fields will be set equal to a function of the
light
fields which transforms as the heavy field does under these
symmetries.
Consequently, the low
energy theory will not exhibit spontaneous symmetry breakdown.
This
argument could fail if $\partial_{\Phi^i} \delta W$ had a large term
which was constant or linear as a function of the $\Phi^i$.

There is a variant of the argument used in \dine\
 to rule out $e^{-{1\over g}}$ contributions to the
superpotential, which also rules out such large terms.  Remember
our
assumption that we are working in a regime in which $e^{- {8\pi^2
\over
g^2}}$ is very small although stringy nonperturbative effects are
large.
The $\Phi^i$ are all charged under the discrete symmetry\foot{An
uncharged VEV would not lead to spontaneous symmetry
breakdown.}, which always
involves a discrete axion shift.  Thus, nonperturbative corrections
to the constant and linear terms in $\partial_{\Phi^i} \delta W$
must
have the form $e^{- R_{0,1} S}$, where $R_{0,1}$ are rational
numbers.
For typical discrete symmetries, assuming that the $\Phi^i$ are
perturbative string states, these rationals are always large enough
that
the new terms can be considered small perturbations of the original
equations.

Notice that this argument proves that the dynamics of the heavy fields does not
break SUSY in the regime where string theory is strongly coupled and the field
theoretic coupling is weak.  SUSY breaking in this regime must then come from
nonperturbative low energy field theory dynamics, and the SUSY breaking scale
will be hierarchically smaller than the string scale.  Note further that we
have proven that the massless spectrum does not change as we move into the
regime of strong string coupling.  (always assuming that the field theoretic
coupling is small).  The quadratic term of the heavy field superpotential is
not significantly altered by the strong dynamics.

There are several loopholes in the above argument which should be mentioned
despite the fact that they appear implausible to us.  First of all, there are
an
infinite number of heavy scalar fields $\Phi^i$ in string theory.  Perhaps this
infinity can alter our naive estimates.  Secondly, the mass of some field can
go to zero despite a large quadratic term in its superpotential, if the Kahler
metric becomes singular.  This would invalidate the assumption of a holomorphic
low energy lagrangian on which our considerations are based.
Finally we note the possibility of exotic soliton states with very small values
of discrete charge, which could alter our estimate of the order of magnitude of
the corrections to the equation which determines the VEV's of heavy fields.
This possibility certainly deserves further study.  It is probably the most
likely way in which our argument could fail.

It is worth while to present an example of the
sort of symmetry which we have in mind.
Consider the Calabi-Yau space based on the quintic
polynmial in $CP^4$ discussed in the text of Green, Schwarz
and Witten,\ref\gsw{M. Green, J. Schwarz and E. Witten, {\it Superstring
Theory}, Cambridge University Press, Cambridge (1987).}.
In this model, there exist, at some points on the moduli
space, a set of $Z_5$ discrete R symmetries.  As the
example is presented in the text, the axion does not transform
under the symmetries.
However, if one includes
Wilson lines,
these symmetries often appear anomalous; the
anomalies can be cancelled by assigning to the axion a
non-linear transformation law of the form:\dineetal
\eqn\axtrans{a \rightarrow a + {2 \pi n \over 5}.}

As an example, consider the point in moduli space with
we can mod out by one of the $Z_5$'s, corresponding
to rotating the coordinates, $Z_a$, of $CP^4 $, by phases:
\eqn\Z{Z_a\rightarrow \alpha^a Z_a}
where $\alpha=e^{2 \pi i /5}$.  This is freely acting;
this means that we don't have to worry about
the appearance of massless particles in twisted sectors
(it leaves a model with $20$ generations).
 This choice leaves over a set
of $R$-symmetries.  For definiteness, consider the
symmetry under which $Z_1 \rightarrow \alpha Z_1$.
Under this symmetry, the gluinos transform by a phase
$\alpha^{-1/2}$.  Now we can
include a Wilson line without breaking this symmetry.
For example, we can include a Wilson line in the ``second"
$E_8$ (the one which is unbroken in the absence of the
Wilson line), described by:
\eqn\wilsoncy{a= {1 \over 5}\left ({1,1,2,0,0,0,0,0}\right ).}
(We are using the notation which is standard in the orbifold
context).  By itself, this choice is not modular invariant, but
this is easily repaired by including a Wilson line in the first
$E_8$ as well.  In the second $E_8$, there are
two unbroken non-Abelian gauge groups.  It is easy to determine
the effects of instantons by simply examining $SU(2)$
subgroups of these.  One finds that instantons
of the first group have four gluino zero modes, while
instantons of the second have $24$.  Thus assigning
to the axion a transformation law
\eqn\funnydelta{a \rightarrow a + {4\pi \over 5}.}
one cancels the anomalies
(This transformation law also cancels the anomalies in the
other $E_8$, for modular invariant choices of the Wilson lines.)

\subsec{The Consequences of Discrete Symmetries}

Having justified the use of discrete symmetries even when the underlying
massive degrees of freedom of string theory
are strongly coupled, we can proceed to use them freely.
Consider the gauge kinetic function of some simple factor of the gauge group.
At tree level this has the form $f_a = {S\over \sqrt{k_a}}$
where $k_a$ is the level of the corresponding Kac-Moody algebra.
The continuous Peccei Quinn symmetry of
perturbative string theory, and the holomorphy of $f_a$ guarantees
(with our definition of renormalization scheme), that the only corrections
to this relation come at one string loop.  Nonperturbatively we cannot rely on
this symmetry
but the discrete gauge symmetries play an analogous role.
In the model discussed in the
previous section, for example, they
guarantee that corrections to $f_a$ beyond
one loop take the form $\delta_{NP} f_a \sim e^{-5S  }(1 +
{\cal O}(e^{- 5 S}))$.  In writing these formulae, we have used
holomorphy of $f_a$, the discrete R symmetry, and the requirement that nothing
blow up at weak coupling.
Our point now is that with a conventional value for the unified coupling in
string theory, the nonperturbative corrections
are extremely small.  By contrast,
we will argue below that stringy corrections to the Kahler potential of the
dilaton can be significant
at these same values of the coupling.  Furthermore, nonperturbative field
theoretic effects like gaugino condensation
have the form $e^{-{2\pi S\over N}}$ for some positive integer $N$.  They are
also much larger than the possible
stringy nonperturbative corrections to the gauge coupling.
Discrete symmetries can thus protect the perturbative string theory prediction
of coupling constant unification even if string theory is strongly coupled at
short distances.

Similar remarks can be made about the superpotential for quarks and leptons.
Perturbative string theory predicts that it is given exactly by its tree level
form.
Discrete symmetries restrict the nonperturbative corrections to be powers of
$e^{- k S}$ where $k$ is
a positive integer determined by the symmetry group.  Again, in order for these
effects to be negligible,
it is sufficient for the effective four dimensional field theory coupling to be
small.  If
this is possible when the string is strongly coupled we will retain these
perturbative
predictions.  The predictions for Yukawa couplings and masses are not so
robust.
These depend on the Kahler potential of the chiral superfields, which we will
argue below may receive
large corrections.  Certain ratios of Yukawa couplings may be independent of
the Kahler potential,
and will therefore be calculable in our framework.  Note that the same sort of
ambiguity
infects the perturbative predictions for Yukawa couplings.  Even in a weakly
coupled
theory where it is calculable,
the Kahler potential depends on the moduli.  Thus, there are no ground state
independent predictions
of couplings in perturbative string theory, except for those combinations of
parameters which
are independent of the Kahler potential.  These are precisely the combinations
that are
calculable in our framework.

Another set of perturbative predictions which cannot be reproduced in our
framework
are results (such as they are) about the structure of soft SUSY breaking terms
in the visible sector.  These depend on the structure of the Kahler potential
in an
essential way.  Furthermore, SUSY breaking can also mitigate the results of the
previous
paragraph about the structure of the superpotential.
It is by now well known
that
SUSY breaking can generate quadratic terms of order $m_{3/2}$ and cubic terms
of order
$m_{3/2}\over M_S$ in the effective superpotential at the gravitino mass scale.
 These can
come from Kahler potential terms in the short distance effective lagrangian,
and are thus
uncalculable in strongly coupled string theory.  Although the effects on
renormalizable
couplings are quite small, they may well be larger than the estimates we made
of nonperturbative
corrections in the previous paragraph.

Finally we note that discrete symmetries may naturally protect the model
independent axion
of string theory from acquiring a large mass.  This might make it a candidate
for solving the
strong CP problem, though such a resolution of the problem will certainly be
fraught with
cosmological difficulties.  We will discuss the axion below, when we take up
the problem
of stabilization of the dilaton in strongly coupled string theory.

\newsec{Stabilization of the Dilaton and Supersymmetry Breaking}

We now come to the topic which forced us to consider strong
coupling string theory in the first place, stabilization of the dilaton
and supersymmetry breaking\foot{It is not at all clear that these two
issues are as closely related in reality as they are in the literature.
Both require violation of perturbative nonrenormalization theorems but
that is the only concrete connection between them.  Indeed, there are
cosmological arguments
 \ref\bkncasas{T. Banks, D.B. Kaplan
and A.E. Nelson, Phys. Rev. {\bf D49} (1994)
779; B. de Carlos, F. Quevedo and
N. Roulet, Phys. Lett. {\bf B318} (447) 1993.}\ and
\modcosm which indicate that the SUSY
breaking scale might be quite different from that at which the dilaton is
stabilized.}.
There are several questions to be
answered here: What are the mechanisms that stabilize the dilaton and break
supersymmetry?  Why is supersymmetry
breaking small if string theory
is strongly coupled?
How is supersymmetry
breaking transmitted to the low energy world? Why is the
unification scale coupling of the effective field theory of the massless
modes small when the underlying string degrees of freedom are strongly coupled?

We begin with the last of these questions.  We have found two alternative
answers to it.  The first, which, as we will
see, appears the most plausible, is based on the observation that
stringy nonperturbative effects
of order $g^{-p}e^{-b\over g}$\shenker\
may contribute
to the Kahler potential of the moduli fields (we have argued above and
in ref. \dine\ that they
cannot contribute to the superpotential or gauge kinetic terms).
If ${g^2 \over 4\pi} \sim {1\over 25}$
then $g \sim .7$.  If $p=0$, then the above
nonperturbative contribution will be as large as a one loop field theoretic
contribution if $b \sim .7  ~\ln 78 = 3.5$.  Thus, it is not implausible that
these effects are significant even when field theory is weakly coupled.

The problem with this argument is that we have very little intuition about the
natural value for the constants $b$ and $p$.  There are two sources of
information about them, exactly soluble low dimensional string theories,
and Wadia's model of a stringy nonperturbative effect as an instanton in
an $SU(2)$ subgroup of a large $N$ gauge theory.   For example, in one
matrix models the $b$ coefficients are all of the form ${2l+1 \over
2l}r_l$, where $l$ is a positive integer and $r_l$ is a (generally
complex) number of modulus less than $1$\ref\gins{P.Ginsparg and J.
Zinn-Justin, in {\it Random Surfaces and Quantum Gravity}, O.Alvarez, E.
Marinari, P. Windey, eds.  Plenum, New York 1991.}
{}.

Wadia's instanton gives us a feeling for
why $b$ need not be a large number like $8\pi^2$.  The action of an $SU(2)$
instanton in a large $N$ gauge theory is ${8\pi^2 \over \lambda^2}N$, where
$\lambda$ is the rescaled coupling.  The $N (= \sqrt{1\over N^2})$ in this
formula plays the role of the string coupling $g$.  The expansion parameter
for the sum of planar diagrams is $\lambda^2 \over 4\pi^2$.  If it is possible
to obtain a critical string theory from large N Yang-Mills theory,
this must be done by tuning the coupling $\lambda$ not to its weak coupling
asymptotically free fixed point, but rather to a finite value where a large
$N$ phase transition takes place.  We would expect this to happen when the
expansion parameter is of order one.  This argument is clearly a general
one and applies to any string theory which is obtained as the limit of a
large $N$ matrix model.

Thus, we might expect that the exponents $b$ in stringy nonperturbative
corrections to the Kahler potential do not contain the ubiquitous geometrical
powers of $\pi$ that appear in all field theoretic instanton calculations.
Perhaps an investigation of the high orders of critical string perturbation
theory can shed further light on this conjecture.  If it is correct, values of
b of order one would be plausible, and nonperturbative string corrections
could indeed be substantial for a four dimensional coupling
${g^2 \over 4\pi}\sim {1\over 25}$.

If one assumes that the string coupling is strong,
there is a second natural way to
explain
the discrepancy in field theory
and string theory coupling strengths.
In the early days of the renaissance of string theory in the 80's, it
was fashionable to use Kaluza-Klein ideas as a bridge between string theory
and ordinary field theory.  Perturbative string theory does not determine the
moduli and it was thought that perhaps they might be determined in such a way
that the internal manifold was larger than the string scale.  It was soon
shown by Kaplunovsky\vadim\ and Dine and Seiberg\ds\ that this
idea is inconsistent with perturbative string theory.  In
superstring theory with large internal manifold, the squared effective coupling
of the four dimensional degrees of freedom is smaller than the squared string
coupling by a factor of the inverse volume of the internal manifold in
string units.  If the string coupling is itself required to be small, then
unless this volume is quite close to one, the predicted unified gauge
coupling will be much too small to be compatible with experiment.

Allowing the string coupling to be large weakens this argument, though
only to a limited extent.
In tree level Kaluza Klein string theory, the $D$
dimensional and $4$ dimensional couplings are related by
\eqn\couplrel{g_4^2 = {g_D^2 \over V_{D-4}}}
where $V_{D-4}$ is the volume of the internal manifold measured in string
units.
When the coupling of the $D$ dimensional theory is large this relation is
corrected by quantum physics.  Unitarity will insure that S matrix elements
in the D dimensional theory are bounded, so it is surely incorrect to imagine
that we can make the volume arbitrarily large for fixed $g_4$ simply by
letting $g_D$ go to $\infty$.  A more reasonable estimate
of the maximum $g_4$ for a given volume is to use the
tree level formula for values of $g_D$ such that one loop corrections in
the $D$ dimensional theory are of order one.
This means $g_D^2 \sim ({4\pi})^{D\over 2} $

One of the attractions of this explanation of the weakness of the coupling is
that
we might be able to link it to the ``observed'' unification of couplings at
$10^{16}$ $GeV$.
In this case we want an internal manifold with scale\foot{We will include
geometrical
factors relevant for a toroidal manifold.  For more general manifolds our
estimates
will change by factors of order 1.} $R \sim {2\pi \over 10^{16}}~ GeV^{-1}$.
It is implausible that
the full $6$ dimensional internal manifold of superstring theory
should be this large.  However, we might consider a manifold where
only $p$ dimensions are larger than the string scale.  If the Wilson lines
which break $E_8 \times E_8$ down to the observed four dimensional symmetry are
wrapped around the large dimensions, then the gauge coupling unification
will take place at the scale $2\pi \over R$.  A little arithmetic shows that
the only plausible choice is $p=1$, corresponding to a six dimensional
``needle''
with length a a few hundred to a thousand times bigger than its width in
the other five compactified dimensions.
This gives a unified four dimensional coupling of order $.18$, for the
circle, which should be compared to the ``observed'' value $.707$.  The
predicted coupling is perhaps a bit small, but our calculations are too
crude to justify rejecting this idea.

J.Polchinski has suggested an orbifold model which realizes this idea,
but also illustrates it's limitations in the strong
coupling context we are considering here.  One
compactifies the heterotic string on the product of three two dimensional tori,
with
complex coordinates $Z_{1,2,3}$, and then mods out by the following
symmetry
\eqn\modout{Z_1 \rightarrow -Z_1 ~~~~Z_{2,3} \rightarrow
i Z_{2,3}}
The transformation has $SU(3)$ holonomy and will give rise to a model with
$N=1$ SUSY in four dimensions.  It will also have chiral fermions.
Note however that we can take the $Z_1$ direction
to be a rectangle, and that we can take one side of this rectangle
arbitrarily large while taking the other of order the string scale.
Thus, if the radius is stabilized at the correct value, this is a model
which might explain the ``data''.

Unfortunately, Polchinski notes,
the Kaluza Klein idea may not be
compatible with our other aim, which
is to stabilize the dilaton.  Above the scale set by $Z_1$,
the theory has five dimensional $N=1$ SUSY and the dilaton is in
a multiplet with a gauge boson.  This determines its Kahler potential
in terms of the analytic gauge kinetic function.  Discrete R symmetries
then restrict the form of its nonperturbative corrections.
It seems that an intermediate Kaluza Klein scale is not compatible
with stabilizing the dilaton, even at strong coupling.
The possible loophole in this argument is provided by twisted states.
These violate $N=2$ SUSY, and in the present context they are strongly
coupled.  It is conceivable that nonperturbative corrections due to
twisted states might rescue this mechanism for explaining the weakness
of four dimensional couplings.

Before ending this subsection let us note that there are many indications
that a string theoretic picture of the world will require more light
particles with standard model quantum numbers than exist in the
supersymmetric standard model.  These are required for example
in the models of \ref\lns{Y. Nir and
N. Seiberg, Phys. Lett. {\bf B309} (1993) 337.},
which attempt to explain the
parameters in the fermion mass matrix, and in many of the known
natural explanations of the absence of flavor changing
neutral currents due to squark exchange.\foot{One particularly
interesting idea to obtain natural flavor conservation
is that of Kaplunovsky and Louis\ref\kl{V. Kaplunovsky
and J. Louis, Phys. Lett. {\bf B306} (1993) 269.}.
However, this scenario
is only viable if the string coupling is genuinely weak.
We will comment on this below.}
If such fields
exist, they will almost certainly change the current picture
of coupling constant unification.  As a consequence, forced to
choose between the Kaluza Klein scenario, which
can explain the ``observed'' coupling unification but perhaps not
the stabilization of the dilaton, and a purely stringy
scenario for strong coupling, whose virtues are exactly opposite,
we opt for the string.  In the next section we argue that
such a scenario indeed has the virtues that we have advertised for it.

\subsec{In Which Supersymmetry Breaking Is Traced to Its Source}

Consider the effective four dimensional lagrangian for the light fields of
string theory at a scale just below the compactification scale.  The arguments
of section II indicate that nonperturbative contributions to the
superpotential of this lagrangian are at most of order $e^{-k S}$
for some positive integer $k$.
As a consequence, {\it stringy non-perturbative effects cannot
be relevant to the problem of supersymmetry breaking in the
real world}, if we assume $S \sim 200 $.   Note that this argument
relies heavily on our assumption of $2 \pi$ periodicity for the
axion; if this is not truly a gauged discrete symmetry of the theory,
the other symmetries we consider here would allow stronger
stringy effects.

By contrast, gaugino condensation in
some factor of the low energy gauge group can give rise to larger terms, of
the form $e^{- {S\over N}}$ for positive integer $N$.  We can however
expect large nonperturbative corrections to the Kahler potential.  Indeed, in
strongly coupled string theory we really do not know how to calculate this
function in the regime of interest.  The flip side of this is that we can
make the Kahler potential responsible for a multitude of sins.  Retribution
will only catch up with us when physicists learn how to calculate
reliably in the strong coupling region

In particular, it is easy to see that the Kahler potential can, with the aid
of a single gaugino condensate, stabilize the dilaton at a SUSY breaking
minimum with zero cosmological constant.    To all orders in
string perturbation theory the Kahler potential is a function of
$S+S^*$.  This is a consequence of the perturbative Peccei-Quinn shift
symmetry of the model independent axion field.
Nonperturbative effects coming from integrating out heavy string modes
will contribute terms of the form $e^{- k S}$ to the Kahler
potential, where $k$ is a multiple of the discrete symmetry index $p$.
Even if $p=1$ this will be smaller than the effects coming from gaugino
condensation which we will discuss below.  It is also much smaller than
stringy nonperturbative effects of the form $e^{-b\sqrt{S+S^*}}$.  Thus
we will discard such terms here, and take the Kahler potential above the
gaugino condensation scale to be a function only of $S+S^*$.

Let us now consider the conventional hidden sector scenario for SUSY
breaking in string theory.  This is based on a gauge group
(``R color'') which commutes
with the standard model group and becomes strong at a scale
$M_R \sim 10^{13.5}$ $GeV$.
R color is taken to be a pure supersymmetric gauge theory, with simple gauge
group.
To all orders in the string loop expansion the gauge kinetic term is given by
\eqn\Rkinterm{\int d^2 \theta S W_{\alpha}^2 + h.c.}
There may be short distance nonperturbative corrections to this, but they are
constrained by symmetries to be very small.  The strongly coupled gauge theory
itself makes a nonperturbative contribution to the superpotential of the
dilaton
below the scale $M_R$.  With our conventions it is exactly
\eqn\npsupot{W_{np} = M_R^3 e^{-  {S / C_A}}}
where $C_A$, the quadratic Casimir of the
adjoint representation, is the coefficient in the
anomaly equation for the gaugino current.
The effective potential of the dilaton superfield is then
\eqn\effpot{V = M_S^3 e^{{-4\pi \over N} y} e^{K(y)}\bigl{(}{[-{2\pi\over
N} + {1\over 4} K^{\prime}(y)]^2 \over {{1\over 4}K^{\prime\prime} (y)}}
- {3\over 4}\bigr{)}}
where $K(y)$, $(y \equiv \ha (S + S^*) = {4\pi\over g^2})$, is the Kahler
potential.
In this equation we have assumed that $M_P = \sqrt{2} M_S$.

Equation \effpot\ has a number of interesting features.  First of all, if
the physical point $y\sim 25$ is in a region where stringy
nonperturbative effects are of order one, then we have no particular
problem in imagining that the potential has a stable minimum with zero
cosmological constant.  This should be contrasted with racetrack models
where one needs at least three independent gaugino condensates and large
numerical coefficients in order to achieve the same results.
Furthermore, in the present case a zero cosmological constant minimum
must break SUSY, since R symmetry is definitely broken.  Again, in
models with complicated superpotentials, this is not necessarily the
case.

The system may have supersymmetric vacua with negative
cosmological constant.  These are not a major worry.  Simple scaling
arguments show that the tunneling amplitude from the zero energy minimum
to one of these states is of order $\exp({- e^{{8\pi \over N}y}})$
per unit
spacetime volume measured in string units. One can further argue
\modcosm\ that the universe will not get trapped in one of these states
at early times.  We want to emphasize that there is nothing in the
formula \effpot\ which requires the existence of negative energy
supersymmetric vacua.  Indeed, for positive potential one can show that
the differential equation which determines $K$ in terms of the potential
always has a solution for finite y.

The scale of SUSY breaking implied by the above potential is $F \sim
{M_S^3 \over M_P} e^{-{4\pi \over C_A}y}$.  Using the ``observed'' value
$y = 25$, $M_P = \sqrt{2} M_S$, this gives $F \sim 2^{-{3\over
2}}e^{-{100\pi \over N}} M_P^2$.  If SUSY breaking is communicated to
the observable sector by gravity, the masses of superpartners of the
ordinary particles will be of order ${F\over M_P}$.  If $N=9$, these
masses come out around $2$ TeV.  Thus, the mechanism described above can
be a plausible description of SUSY breaking in the real world.

\subsec{The String Axion as the Axion of QCD}

The final feature of this potential which we want to point out is its
independence of the axion field.
The renormalizable terms in the
lagrangian have an accidental anomalous $U(1)$ R-symmetry.  When
combined with the shift symmetry of the axion, we obtain an anomaly free
continuous R symmetry.  This symmetry is broken already
in perturbation theory by higher dimension operators.
However, in the presence of discrete symmetries, the
leading operators which violate the symmetry
may be of quite high dimension.
To understand the size of PQ symmetry-violating effects,
consider first operators involving only hidden sector gauginos.
In the case of a $Z_5$ R symmetry, for example, the
leading symmetry-breaking operator is $(\lambda \lambda)^5$,
which has dimension $15$.  We might then expect the hidden-sector
contribution to the axion mass to be of order $\Lambda^{17}/M_p^{15}$,
which is smaller than $10^{-9} f_{\pi} m_{\pi}$ for
$\Lambda < 10^{15} GeV$.  Other contributions which might
arise due to symmetry-violating couplings to, e.g., light fields,
can be shown to be even further suppressed.

\newsec{\bf Summary and Conclusions}

String theory, if it describes nature, is almost certainly
strongly coupled.  There is little hope for understanding
strongly coupled string theory in the near future,
so it would seem that there is no chance of establishing
the truth (or falsehood) of string theory by making
predictions for low energy theory.  We have seen here, though,
that this is not the case.  By making certain assumptions,
one can make a limited but quite well-defined set of predictions.
These assumptions, that the cosmological constant
vanishes at the minimum, that at the minimum the dilaton
vev is large, and that the true minimum is connected
to a perturbative ground state by varying the dilaton
are all very strong, but they
are also likely to be true if string theory describes nature.
Moreover, this is probably the best one can do.

It is useful at this stage to summarize the phenomenology
of the strong coupling theory, and compare it with
discussions of weak coupling string theory.
There are several
which are generic, some of which we have already mentioned:
\item{1.}
Existance of a hierarchy between the supersymmetry
breaking scale and the string scale.  A priori, we might
have imagined that if string theory is strongly
coupled, supersymmetry breaking should occur
at the string scale.  However, we have seen that
the assumption of small gauge couplings, as observed
in nature, implies that the superpotential is very
small.  Indeed we have argued that stringy non-perturbative
effects can not give phenomenologically interesting
supersymmetry breaking; this must arise from
effects visible in the low energy theory.  These statements
relied on our argument that the $2 \pi$ periodicity of the
axion is exact; if this is not the case, it is possible for
stringy-non-perturbative effects to play a role comparable to
gluino condensation.
\item{2.}
The light spectrum:  As we have already noted, in this
framework, it follows that the low energy spectrum
is the same as that at weak coupling.
\item{3.}
Gauge coupling unification:  The gauge couplings are unified.
We have already discussed how the function, $f$, in a suitable
scheme, is not renormalized beyond one loop.  However,
this does not mean that we can compute exactly the
coupling unification in strong coupling.   As discussed in ref.
\ref\yuri{M. Dine and Yuri Shirman, SCIPP preprint
SCIPP 94/11 (1994).}, even in the Wilsonian effective action,
it is necessarily to carefully choose the cutoffs if one
is to maintain holomorphy of $f$.  The appropriate cutoffs
must be determined order by order in perturbation theory.
In strong coupling, one might expect these cutoffs to shift
by amounts of order one (this is similar to the
expected shifts of thresholds).  Thus the prediction of coupling
constant unification is valid only to order one shifts of
the unification scale.  Of course, one might hope for shifts
of factors of $100$ or so, but this does not seem terribly likely.
\item{4.}
Grand unified prediction for gaugino masses:  There
is at least one generic prediction for the structure
of soft breaking terms.  This again arises from the symmetry
constraints on the function $f$ which describes the gauge
couplings.  The leading term in this coupling is the tree
level dilaton term; at one loop, moduli couplings may appear.
At the unification
scale, provided the dilaton F-term is comparable to or larger
than the moduli $F$-terms, the dominant contribution will
be from the universal dilaton term, so the gaugino masses will
be equal at this scale; at lower scales, as is well-known, they
then go as ratios of the appropriate gauge couplings.
\item{5.}
Non-renormalization of the matter superpotential:  the
superpotential of the matter fields is only corrected by
exponentially small effects from its tree level value
in this picture.  In any given compactification, this means
that there are some number of predictions, for example,
of Yukawa couplings.  As we have stressed, this is similar
to the situation in perturbative string theory, if one does
not know the expectation values of the moduli.

It is perhaps useful to mention a few type of predictions
which have been discussed in the literature which are
not expected to hold, in any generic sense, in this strong
coupling picture.  These are statements which require the
corrections to the Kahler potential be small, which, by
assumption, is not the case here.
Perhaps the most interesting discussion of soft breaking in
string theory is that due to
Kaplunovsky and Louis\kl, who have pointed out that there is
a circumstance in string theory in which one might
expect squark degeneracy at the high scale, and
corresponding suppression of flavor-changing processes.
If the dilaton auxiliary field is the principle source
of supersymmetry breaking, they note that, because
of the universal character of tree-level dilaton couplings,
the leading contributions to squark and slepton masses
are identical.  This is a quite appealing result; it is the
only rationale which has every been offered for universal
squark and slepton masses at the Planck mass.  It is also
interesting, in that one-loop effects probably give corrections
at best just barely consistent with the limits from the
$K$-$\bar K$ system.  This scheme, however, will not
operate in any generic fashion in strongly coupled strings.
While it is possible in
this scheme to obtain ``dilaton domination" in strongly coupled string theory
(e.g. as a consequence of the
action of symmetries on the moduli fields),
there is no reason to expect that the full Kahler
potential maintains the universality of the tree level result.
Already in perturbation theory, there are corrections
which do not respect
this universality.  Thus the problem of flavor changing
neutral currents will have to be solved in some other way,
perhaps using a flavor symmetry
along the lines of refs. \ref\dkl{M. Dine, A. Kagan
and R.G. Leigh, Phys. Rev. {\bf D48} (1993) 4269.}
and \lns, or through
renormalization group effects as in ref. \ref\dks{M.
Dine, A. Kagan and S. Samuel, Phys. Lett. {\bf B243} (1990)
250.}

\centerline{\bf Acknowledgements}

It is a pleasure to thank G. Moore, J. Polchinski, N. Seiberg,
S. Shenker and E. Witten
for helpful comments and suggestions.  The work of M. Dine was supported in
part
by the U.S. Department of Energy.  The work of T. Banks was supported in
part by the Department of Energy under grant$\# DE-FG0590ER40559$.

\bigskip
\bigskip
\centerline{\bf Appendix A.  Discrete Symmetries and Their
Breaking}
\centerline{\bf  In Supersymmetric QCD}

In this paper we have used spontaneously broken discrete symmetries to tightly
constrain the form of the low energy effective action.  We have argued
that this is permissible because the symmetry breaking is do to a light
field, the axion.
There are, in fact, a set of well-studied field theories which
exhibit this sort of behavior:  supersymmetric QCD with gauge
group $SU(N)$ and $N_f$ flavors, where $N_f<N$.    By $N_f$
flavors, here, one means a set of $2N_f$ fields, $Q_f$ and
$\bar Q_f$, transforming
in the $N$ and $\bar N$ representations, respectively.
\foot{The treatment of ref. \ref\ads{I. Affleck,
M. Dine and N. Seiberg, Nucl. Phys. {\bf B241} (1984) 493.}
which we follow here,
most closely parralels the structure observed in string
theory.  Other treatments can be found in ref. \ref\otherqcd
{A. Amati, K. Konishi, V. Meurice, G.C. Rossi and G. Veneziano,
Phys. Rep. {\bf 162} 1988) 169.}.}

Consider, first, the case where the ``quarks" are massless.
These theories have, at the classical level, a continuum of
ground states, quite analogous to those of string theory.  In
these, up to gauge and flavor transformations, the general
flat direction has the form
\eqn\flatdirections{Q = \bar Q = \left ( \matrix{v_1 & 0  & \dots \cr 0 & v_2
&\dots
 \cr \dots  \cr 0 & \dots & v_{N_f} } \right ).}
In these directions, the gauge symmetry is broken to $SU(N-N_f)$.
The corresponding gauge fields gain mass of order $gv$.   To
understand the vacuum structure of the theory, one wants
to construct an effective action describing the low
energy theory in these flat directions.  This action is highly
constrained by the symmetries.  These include an $SU(N_f)_L
\times SU(N_f)_R$ symmetry, a vector $U(1)$, and a
non-anomalous $R$ symmetry under which
\eqn\rsymmetry{\lambda \rightarrow e^{i \alpha} \lambda
{}~~~~~~Q \rightarrow e^{i{N-N_f \over N_f} \alpha} Q
{}~~~~~~\bar Q \rightarrow e^{i{N-N_f \over N_f} \alpha} \bar Q.}
These symmetries determine the form of the superpotential uniquely; it can
be written in terms of a chiral field,
$\Phi=det(Q_f \bar Q_{f^{\prime} })$.
\eqn\wnp{W_{np} = {A \Lambda^{3N-N_f \over N-N_f}
 \over \Phi^{1 \over N-N_f}}}
where $\Lambda$ is the scale parameter of the theory.
That such a superpotential is in fact produced has long
since been verified.

Now suppose we add a small mass term to this theory
(for convenience taken to be $SU(N_f)$ symmetric),
\eqn\wmass{W_o = mQ \bar Q.}
In this case, the continuous $R$ symmetry
described above is explicitly broken, but there
is still a non-anomalous discrete symmetry (i.e. a symmetry
unbroken by instantons) under which
\eqn\discretesymm{\lambda \rightarrow e^{2 \pi i \over N}\lambda~~~~~
Q \rightarrow e^{2 \pi i \over N}Q~~~~~
\bar Q \rightarrow e^{2 \pi i \over N} \bar Q.}
However, unlike for the case of the continuous symmetry,
this discrete symmetry is not respected by the
non-perturbative superpotential, $W_{np}$, except
when $N_f = N-1$.  It is also interesting to note that,
except, again for this special number of flavors, $W_{np}$
has branch cuts.

To understand these phenomena, let us return to the
massless theory and look
more closely at the dynamics in the flat directions.
When $N_f < N-1$, there is an unbroken gauge
symmetry in the flat directions, $SU(N-N_f)$.  The
light particle content consists of the gauge bosons
and gauginos of this gauge group, as well as the Goldstone
particles associated with broken global symmetries
and their superpartners.  The $SU(N-N_f)$ gauge
theory becomes strong
at some scale, $\Lambda_{N-N_f}$, and is believed to produce a
(supersymmetric)
set of bound states with masses of order $\Lambda_{N-N_f}$.
In addition, it is believed that gluino condensation occurs.

Below the scale $\Lambda_{N-N_f}$, one has only the Goldstone supermultiplets;
$W_{np}$ represents a superpotential appropriate to
their interactions.  To understand how this arises, it is
convenient to look at an $SU(N_f)$-symmetric flat direction,
$v_1=\dots =v_{N_f}$,
and parameterize the fields in this direction such that
\eqn\axiondefined{\Phi = \rho e^{ia},}
where $\rho$ is a massless field with
$<\rho>=v^{2 N_f}$.  Under the continuous
$U(1)_R$, $a \rightarrow a + \alpha(N-N_f)$.  In the
theory below the scale $v$, it is easy to check that
triangle diagrams generate a coupling
\eqn\acoupling{a {1 \over 32 \pi^2} F \tilde F,}
where the gauge fields are those of the $SU(N-N_f)$.  This
coupling insures that the theory at scales larger than
$\Lambda_{N-N_f}$ respects the (non-linearly realized)
$R$ symmetry.  Its supersymmetric expression is
\eqn\lnphiterm{{1 \over 32 \pi^2} \int d^2 \theta~ ln (\Phi) W_{\alpha}^2.}
It is perhaps worth noting that this coupling, which is obtained by
integrating out massive particles, is holomorphic.
The gluino condensate then gives rise to an $F$-term for
$\Phi$.  This is the origin of the non-perturbative superpotential.
In order to understand how this $F$ term depends on the fields,
note that
\eqn\lambdalambda{<\lambda \lambda> =e^{2 \pi i {n \over N-N_f}}
e^{i{ a \over N-N_f}}
\Lambda_{N-N_f}^3.}
The first term represents the fact that the pure $SU(N-N_f)$
gauge theory has a $Z_{N-N_f}$ symmetry, broken by
the condensate; n is an integer which runs from $1$ to $N-N_f$.
The second term describes the dependence of the condensate
on the axion (which can be obtained from standard anomaly
arguments, as in QCD), and the last term follows from dimensional
analysis.  Finally,
\eqn\lambdannf{\Lambda_{N-N_f}^3 \sim v^{1/(N-N_f)}.}
This gives precisely the dependence on $v$ and $a$
expected from $W_{np}$.

We are now in a position to answer the various questions
we raised earlier.  First, we can understanding the
appearance of branches of the superpotential; these are
associated with the different choices of the phase of the
condensate labeled by the integer $n$.  The condensate breaks
the approximate $Z_{N-N_f}$ symmetry of the intermediate
energy theory.
We can also answer what happens in the presence of quark
mass terms to the discrete $Z_N$ symmetry of the full theory.
$a$ transforms non-linearly under this symmetry, but
it is also broken by the condensate.  Indeed,
under this symmetry, $<\lambda \lambda>$ is not invariant;
it transforms as $e^{2 \pi i \over N}$.  At scales below
$\Lambda_{N-N_F}$, the $\lambda$'s are to
be thought of as massive fields.  Integrating them out,
we obtain the non-perturbative superpotential of the low
energy theory, $W_{np}$, which no longer need respect
the symmetry.  Indeed, from a ``microscopic perspective,"
the coefficient,
$A$, in $W_{np}$ transforms like $<\lambda \lambda>$,
and the full superpotential transforms, as
it should, by $e^{4 \pi i\over N}$.
This is as we would expect: in the low energy theory, phases
appear corresponding to the discrete choices of phases in
massive fields; these phases can be compensated by
performing the discrete transformation on the light fields.
Note that, in the
theory with zero quark mass,
there is no such effect; the dependence
of the condensate on the Goldstone boson is fixed
by the symmetry, and the symmetry is realized in the
lagrangian at the low scale.  It is interesting
to understand the connection of the $Z_{N-N_f}$
symmetry and the $Z_N$ symmetry.  Under
this symmetry,
\eqn\zminumnf{<\lambda \lambda> \rightarrow \Lambda^3_{N-N_f}
e^{i{a \over N_f-N}} e^{2 \pi i n \over N-N_f} <\lambda \lambda>.
}
In other words, written in terms of the transformed axion field, this
is a $Z_{N-N_f}$ transformation.

So we see that it is the gluino
condensation in the intermediate scale theory
which accounts for the lack of invariance of the low
energy theory under the discrete symmetry.
In other words, in the theory below
the scale $v$, not just $a$ but also $\lambda$ transforms
under the symmetry.  This symmetry is still present in the
theory at scales above $\Lambda_{N-N_f}$.  Below  this scale,
the dynamics of $\lambda$ further breaks
the symmetry, and the theory at lower energies shows no relic
of the symmetry (except for the existance of the branches).

To futher verify this picture, consider finally the case
that $N_f=N-1$.  In this instance, in the flat directions
there is no unbroken gauge group; only $a$, among the
light fields, transforms under the discrete symmetry.
So the low energy effective action must respect the
symmetry.  Indeed, the non-perturbative
superpotential $W_{np}$ does respect it.

These models appear quite analogous to string theory.  At energies below
the string scale, one has a discrete symmetry; at least perturbatively,
none of the very massive fields break it.  Any breaking should be
due to the light fields, $a$, and perhaps gluinos or other
fields.  This breaking should be understandable in the low energy
(below the string scale) theory.

\listrefs
\bye
\end